# Mechanics of Schrodinger mechanics

**Valery P. Dmitriyev**

Lomonosov University,

P. O. Box 160, Moscow 117574, Russia

dmitr@cc.nifhi.ac.ru

25 November 2003

Small perturbations of ideal turbulence obey the Schrodinger equation. Microscopically, the perturbation of turbulence corresponds to formation of small amplitude helices on straight vortex filaments. A helix behaves in the vorticity field of the fluid as a spin particle in the Stern-Gerlach experiment. Taking into account elastic properties of the filament leads to the Klein-Gordon equation.

## Introduction

Schrodinger equation was conceived by the author as describing some deterministic wave similar to the wave in a medium. The nonlinear Schrodinger equation is now suggested as a rather universal model of turbulence (see e.g. [1, 2]). So, we may conclude that small perturbations of ideal turbulence provide us with continuum mechanics realization of Schrodinger mechanics.

The intrinsic feature of this mechanism is given by dynamics of a vortex filament that is recognized as a constitutive element of turbulence [3]. Below we will try to derive the equation of motion of a turbulent medium, assuming that its structure element is a straight vortex filament. We will take a string as a generalized model of a vortex filament. A string evolves in time by certain laws. We will consider a string that only slightly deviates from the rectilinear configuration. Assuming that adjacent parts of the string interact with each other as two vortices we will show that the its evolution obeys the Schrodinger equation. In this event, the string takes the configuration of a small amplitude helix. Supposing additionally that the string can be stretched elastically we come to the Klein-Gordon equation. We will show also that the helices behave in the vorticity field of the fluid just as spin particles in the Stern-Gerlach experiment.

## A string

Consider a spatial curve. In general, its position in space and time is specified by the radius vector

$$\mathbf{r}(\mathbf{x}, \mathbf{y}, \mathbf{z}, t) . \qquad (1)$$

We are interested in the case when the curve is only a slight deviation from a straight line. If the latter is directed along the $\mathbf{x}$ axis then we may write

$$\mathbf{r} = x\mathbf{i}_1 + y(\mathbf{x}, t)\mathbf{i}_2 + z(\mathbf{x}, t)\mathbf{i}_3 , \qquad (2)$$

and the space-time position of the curve is specified by the plane vector

$$\mathbf{s} = y(x,t)\mathbf{i}_2 + z(\mathbf{x},t)\mathbf{i}_3. \tag{3}$$

This defines the frame of reference for an object further referred to as a string.

## An elastic string

The behavior of a string that can be stretched elastically moving along the **y** and **z** axes is known to obey the d'Alembert equation

$$\frac{\partial^2 \mathbf{s}}{\partial t^2} = c^2 \frac{\partial^2 \mathbf{s}}{\partial x^2}. \tag{4}$$

A solution to Eq. (4) is given by two independent functions:

$$\mathbf{s} = y(x-ct)\mathbf{i}_2 + z(x-ct)\mathbf{i}_3. \tag{5}$$

Eq. (5) specifies a hump that moves along the **x** axis with the constant velocity $c$.

## A vortex filament

A vortex generates in the fluid a circular velocity field. So, two vortices will rotate each other. The same is true for two vortices that are two different parts of one and the same vortex filament. This is just the cause why a curved vortex filament evolves. Assuming that only adjacent parts of the filament influence each other we come to the following equation

$$\frac{\partial \mathbf{r}}{\partial t} = \nu \frac{\partial \mathbf{r}}{\partial l} \times \frac{\partial^2 \mathbf{r}}{\partial l^2}, \tag{6}$$

where $l$ is the distance along the filament, $\nu$ he coefficient of local self-induction [4, 5] (see Fig.1).

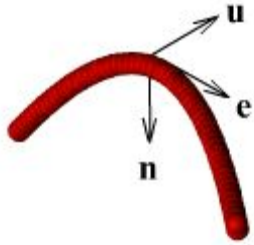

Fig.1
The drift $\mathbf{u} = \partial \mathbf{r}/\partial t$ of a bent vortex filament. Other vectors show the direction
$\mathbf{e} = \partial \mathbf{r}/\partial l$ of the filament's vorticity and its curvature $\kappa \mathbf{n} = \partial^2 \mathbf{r}/\partial l^2$.

We may assume for small deviations from the straight line:

$$dl = dx. \tag{7}$$

That gives for Eq. (6)

$$\frac{\partial \mathbf{r}}{\partial t} = \nu \frac{\partial \mathbf{r}}{\partial x} \times \frac{\partial^2 \mathbf{r}}{\partial x^2}. \tag{8}$$

Substituting (2) into Eq. (8) and neglecting quadratic terms we get two scalar equations

$$\frac{\partial y}{\partial t} = -\nu \frac{\partial^2 z}{\partial x^2}, \tag{9}$$

$$\frac{\partial z}{\partial t} = \nu \frac{\partial^2 y}{\partial x^2}. \tag{10}$$

A solution to the system of equations (9) and (10) is given by

$$y = a\cos\left(\frac{x}{b} - \frac{v}{b^2}t\right), \tag{11}$$

$$z = a\sin\left(\frac{x}{b} - \frac{v}{b^2}t\right). \tag{12}$$

Eqs. (11) and (12) describe a helix with the amplitude $a$ and the thread, or pitch, $b$. The helix rotates counter-clockwise around the **x** axis looking in the direction of the **x** axis with the angle velocity

$$\omega = \frac{v}{b^2}. \tag{13}$$

(see Fig.2).

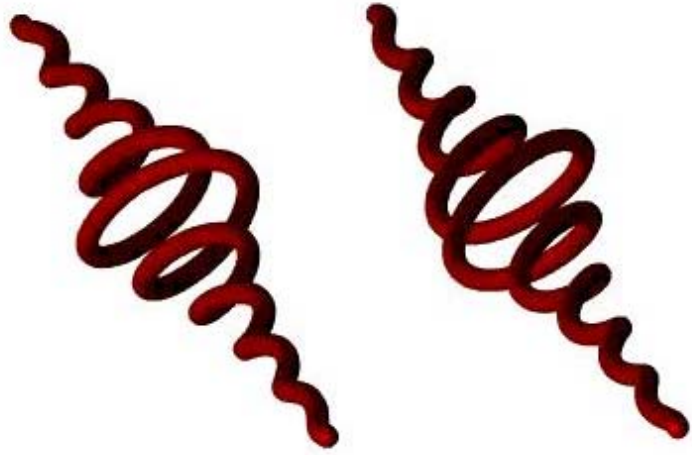

Fig.2

Wave packets comprised of left-hand screw helices (left) and right-hand screw helices (right).

The system of linear equations (9) and (10) is valid only provided that

$$a \ll |b|. \tag{14}$$

This system can be united into one equation for the complex-valued function

$$\Phi = y + iz. \tag{15}$$

To this end, we multiply Eq. (10) by $i$ and substitute $-1 = i \cdot i$ into Eq. (9). Summing resulting equations we get

$$\frac{\partial \Phi}{\partial t} = iv\frac{\partial^2 \Phi}{\partial x^2}. \tag{16}$$

Inserting Eqs. (11) and (12) into (15) gives the complex-valued representation of the helix

$$\Phi = a\,\exp[i(\tau x - v\tau^2 t)] \tag{17}$$

where

$$\tau = \frac{1}{b} \tag{18}$$

is the torsion of the helix.

It can be shown [6] that large deviations of the string from the straight line obey a nonlinear Schrodinger equation. The latter describes a soliton on the vortex filament that has the form of a loop. A curvilinear configuration of a nonstretchable filament can be obtained only by adding to the straight vortex filament a filament's segment. In the case under consideration the length of the additional segment equals to

$$2a.  \tag{19}$$

Alternatively the soliton can be seen as a wave packet comprised of the helices described by Eqs. (11) and (12) (for details see [5]).

The nonlinear analysis shows that the helix on the vortex filament should be properly described by the equation

$$\frac{\partial \Psi}{\partial t} = i\nu \frac{\partial^2 \Psi}{\partial x^2} \tag{20}$$

that is simply a time derivative of Eq. (16):

$$\Psi = \frac{\partial \Phi}{\partial t}. \tag{21}$$

So, the wave function has the meaning of the velocity which the helix rotates around the screw axis.

## An elastic vortex filament

The elastic stretching of the vortex filament causing a move in the direction of **y** and **z** axis can be also taken into account. In order to describe this we combine Eqs. (4) and (16) into a single equation. To this end let us rewrite Eq. (4) to a complex-valued form:

$$\frac{\partial^2 \Phi}{\partial t^2} = c^2 \frac{\partial^2 \Phi}{\partial x^2} \tag{22}$$

where $\Phi$ is given by Eq. (15). Next, we will take advantage of the fact that a helix provides also a particular solution to Eq. (22). The left-hand part of Eq. (4), or (22) has the meaning of the acceleration. In order to find an addition to the acceleration due to self-induction of a vortex filament we take twice the derivative of (17) in respect to time. That gives

$$\frac{\partial^2 \Phi}{\partial t^2} = -\nu^2 \tau^4 \Phi. \tag{23}$$

Adding the right-hand part of Eq. (23) to the right-hand part of Eq. (22) we get the equation

$$\frac{\partial^2 \Phi}{\partial t^2} = c^2 \frac{\partial^2 \Phi}{\partial x^2} - \nu^2 \tau^4 \Phi \tag{24}$$

that describes both the elastic stretching and the self-induction of the string.

## The spin

We may distinguish two kinds of helices that differ in the sign of the torsion (18). The right-hand screw helix (Fig.2, right) is described by Eq. (17) with $\tau > 0$. The left-hand screw helix (Fig.2, left) is described by (17) with $\tau < 0$. As we can see readily from Fig.1, both kinds of helices rotate in the same direction, the moment of the helix rotation being directed counter to the vorticity vector of the rectilinear part of the filament (that was chosen in (17) so as to coincide in direction with the **x** axis).

Let us compose from the helices of one sign of the torsion $\tau$ a wave packet such as it is done with harmonic waves in quantum mechanics. The resulting composition will move along the vortex filament with the group velocity

$$\upsilon = 2\nu\tau. \tag{25}$$

Eq. (25) says that the right-hand screw helix travels in the direction that the filament's vorticity points to. The left-hand screw helix goes in the opposite direction. In order that these two kinds of helices move in one and the same direction they should be resided on antiparallel vortex filaments (Fig.3). Then these two helices will have opposite rotational moments **μ**.

Let us impose on this system a fluid vorticity field $\operatorname{curl}\langle\mathbf{u}\rangle$, where $\langle\mathbf{u}\rangle$ is the average velocity of the fluid. A helix will interact with the vorticity field by the common formula for the energy of interaction

$$-\boldsymbol{\mu}\cdot\operatorname{curl}\langle\mathbf{u}\rangle. \qquad (26)$$

If the field is inhomogeneous the force will act upon a helix. Let we have an ideal fluid pierced in all directions by straight vortex filaments. Let two kinds of helices move in the direction of the **x** axis. Let a fluid vorticity field exists that grows in the direction of the **y** axis. In order to diminish the energy (26) of interaction with the vorticity field a rotational moment $\boldsymbol{\mu}$ must be oriented along the vector $\operatorname{curl}\langle\mathbf{u}\rangle$. A helix can be turned around by its torsion axe if only it will jump over to another filament. This must be an adjacent filament with similar but slightly different orientation.

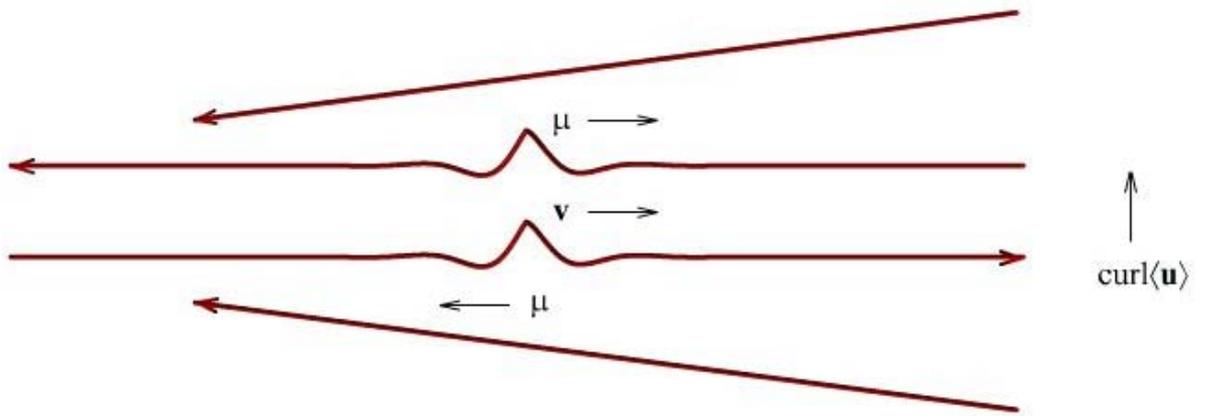

Fig.3

A right-hand screw helix (bottom) and left-screw helix (top) traveling from the left to the right in the vortex sponge through an inhomogeneous field of fluid vorticity $\operatorname{curl}\langle\mathbf{u}\rangle$. Arrows on the filaments indicate the direction of their vorticity, **v** shows the direction of the translational motion and $\boldsymbol{\mu}$ the rotational moments of the helices.

In the result, the left-hand screw helix (Fig.3, top) will deviate in the direction of the **y** axis. The right-hand screw helix (Fig.3, bottom) will deviate in the opposite direction. The helices will deflect in the inhomogeneous vorticity field $\operatorname{curl}\langle\mathbf{u}\rangle$ such as two spin particles do in the inhomogeneous magnetic field $\mathbf{H}=\operatorname{curl}\mathbf{A}$. This reproduces conditions of the Stern-Gerlach experiment.

## Quantum mechanics analogy

Taking for the particle's mass

$$m = \varsigma\, a, \qquad (27)$$

where $\varsigma$ stands for linear density of the fluid along the filament, and defining

$$\hbar = 2v\varsigma a, \qquad (28)$$

we come in Eq. (20) to the standard form of the Schrodinger equation.
The energy integral of the soliton on a vortex filament was shown [7] to be

$$E = 2\varsigma a v^2 \tau^2. \qquad (29)$$

Assuming in Eq. (29) $E = mc^2$ and using Eqs. (27) and (28) we come in Eq. (24) to the standard form of the Klein-Gordon equation.

# Conclusion

Thus, we find an elementary mechanical object that obeys the main equations of quantum mechanics. This object is a vortex filament and it is essentially one-dimensional. To extend this dynamics to three dimensions is a nontrivial task and it can not be done analytically. For, here we deal with the system of infinite degrees of freedom. However, the three-dimensional Schrodinger equation is now widely used in order to describe ideal turbulence [2]. Its validity is confirmed both by numerical calculations and experimentally. On the other side, that the system perceived on a large scale as a turbulent fluid has the vortical microscopic structure is also nowadays widely recognized. This enables us to draw the following conclusion.

An ideal fluid pierced in all directions by straight vortex filaments represents a substratum for deterministic Schrodinger field. The vortical structure with elastically stretchable filaments supports as well the Klein-Gordon field.

**P.S.**

# Quantum mechanics demystified

There are two features that should be added to the model in order to get quantum mechanics. They are the wave function collapse and stochasticity. The latter is inherent in the very notion of turbulence. The instantaneity of the collapse of the wave function is ensured by that the ideal fluid is incompressible.

We see the substratum as a vortex sponge. In the infinity it may be a tangle rolled from a single vortex tube. On a moderate scale it is seen as a fluid pierced by straight vortex tubes, or filaments.

A particle corresponds to an additional segment of the vortex tube. Thanks to this addition the tube deviates from the rectilinear configuration. By hydrodynamic relations for the energy [5] the loop on the vortex filament thus formed tends to have a specific size. In general, a particle is seen as some nonlinear vortex configuration formed on the linear vortex sponge substratum.

In the stochastic environs the loop is thermalized splitting into small helices that evolve over the filaments according to Schrodinger equation. This looks as a kind of diffusion and proceeds with a finite velocity. Mathematically, the dispersing of a soliton into asymptotic helices corresponds to the Fourier decomposition of a function into the sum of harmonics. In order that the model will be consistent in the definition of particle's mass, the concept of elementary helix should be introduced. So that in the definition of the Planck constant (28), $a$ must be understood as the length of an elementary helix [5]:

$$a_0 = \frac{2\nu}{c}. \tag{30}$$

Next, let a drop of the pressure occurs at some point of the fluid. This models the measurement of the particle. The drop of the pressure destroys the dispersion cloud causing the gathering of the redundant emptiness into the point. The process runs as a shock wave and in the form of avalanche. The speed of the longitudinal wave in the vortex sponge formed in an incompressible medium can be infinitely high. The occurence of the two distinct time scales for the processes in a stirred up fluid can be clearly observed in experiments [8]. There are two kinds of waves spreading over vortex tube: the slow spiral wave and very fast varicose (area-varying) wave.

So, hidden parameters of quantum mechanics may be a turbulent ideal fluid with the inclusion of some empty space and having the vortex sponge microscopic structure. The "philosophical" ban imposed on hidden parameters by some influential politicians led to that quantum mechanics was mystified. Most popular mystifications are the so called quantum logic, diffusion with an imaginary diffusion coefficient and evolution in imaginary time.

The question of supposedly a violation of the logic in many-slit one-particle diffraction is resolved in the aether model by that we see a particle as an inclusion of empty space into a medium. Initially it may be a single cavity. Then it is dispersed in the medium. And after the hit into the screen

7it is again collected into a bubble. The spreading of the emptiness over the medium proceeds with a finite velocity. Its gathering into a localized cavity happens instantaneously.

Concerning the speculations about imaginary time and diffusion coefficient, in order to expose them we must return to the law (6) of the motion of a vortex filament. Let us compare it with the law of motion of an elastic string. Firstly, take notice that Eq. (6) defines the dependence of the velocity of the filament on its configuration. That is why we have the first derivative with respect to time in the left-hand part of the Schrodinger equation (16). Whereas the law of motion of an elastic string expresses the dependence of the acceleration on the configuration of the filament. So, we have the second derivative with respect to time in the left-hand part of the d'Alembert equation (22).

The linear Schrodinger equation is concerned with an asymptotic solution of Eq. (6) that is a helix with the amplitude small comparing with the pitch as it is shown in relation (14). In this configuration the filament's vorticity is actually parallel to the direction of the **x** axis:

$$\frac{\partial \mathbf{r}}{\partial l} = \mathbf{i}_1 \tag{31}$$

where Eqs. (7) and (2) were used. The curvature vector $\kappa \mathbf{n} = \partial^2 \mathbf{r}/\partial l^2$ of the asymptotic helix is directed to the **x** axis:

$$\begin{aligned}\frac{\partial^2 \mathbf{r}}{\partial l^2} &= \frac{\partial^2 \mathbf{r}}{\partial x^2} = \frac{\partial^2 y}{\partial x^2}\mathbf{i}_2 + \frac{\partial^2 z}{\partial x^2}\mathbf{i}_3 \\ &= -a\tau^2\left[\cos(\tau x - \omega t)\mathbf{i}_2 + \sin(\tau x - \omega t)\mathbf{i}_3\right] \\ &= -\tau^2\left[y\mathbf{i}_2 + z\mathbf{i}_3\right]\end{aligned} \tag{32}$$

where Eqs. (7), (2) and (9), (10) with (13), (18) were used. According to Eq. (6) with (31), in order to obtain from the curvature vector the velocity of the filament we must rotate it by 90° counter-clockwise around the **x** axis (see Fig.1). In complex-valued notations this means that we must multiply the curvature $\partial^2 \Phi/\partial x^2$ by the imaginary unit $i$ where $\Phi$ is given by (15). This is the only "physical meaning" of complex values occurred in the Schrodinger equation.